\documentclass[conference]{IEEEtran}
\IEEEoverridecommandlockouts
\usepackage{cite}
\usepackage{amsmath,amssymb,amsfonts}
\usepackage{graphicx}
\usepackage{booktabs}
\usepackage{subcaption}
\usepackage{textcomp}
\usepackage{xcolor}
\usepackage{multirow}
\usepackage{algorithm}
\usepackage{algpseudocode}
\usepackage{url}
\usepackage{lineno,hyperref}
\def\BibTeX{{\rm B\kern-.05em{\sc i\kern-.025em b}\kern-.08em
    T\kern-.1667em\lower.7ex\hbox{E}\kern-.125emX}}

\begin{document}

\title{Compass: Dissecting Communication and Computation Operators for Efficient LLM Training}

\author{
\IEEEauthorblockN{Guangyu Xiang\IEEEauthorrefmark{2}, 
Lin Zhang\IEEEauthorrefmark{3}\IEEEauthorrefmark{1}, 
Haoxuan Yu\IEEEauthorrefmark{4}, 
Xinglin Pan\IEEEauthorrefmark{2}, 
Shaohuai Shi\IEEEauthorrefmark{4},  
Xiaowen Chu\IEEEauthorrefmark{2}\IEEEauthorrefmark{3}\IEEEauthorrefmark{1}\thanks{* Corresponding authors.}\\}

\IEEEauthorblockA{
\IEEEauthorrefmark{2}Data Science and Analytics Thrust, The Hong Kong University of Science and Technology (Guangzhou)\\
\IEEEauthorrefmark{3}Department of Computer Science and Engineering, The Hong Kong University of Science and Technology\\
\IEEEauthorrefmark{4}School of Computer Science and Technology, Harbin Institute of Technology, Shenzhen\\
gxiang190@connect.hkust-gz.edu.cn, lzhangbv@connect.ust.hk, 23s151042@stu.hit.edu.cn, \\xpan413@connect.hkust-gz.edu.cn, shaohuais@hit.edu.cn, xwchu@ust.hk}
}

\maketitle

\begingroup
\renewcommand\thefootnote{}
\footnotetext{© 2026 IEEE. Personal use of this material is permitted. Permission from IEEE must be obtained for all other uses, in any current or future media, including reprinting/republishing this material for advertising or promotional purposes, creating new collective works, for resale or redistribution to servers or lists, or reuse of any copyrighted component of this work in other works.}
\endgroup

\begin{abstract}

Overlapping communication and computation operators is a common practice to hide communication overheads, accelerating large language models (LLMs) training on GPU clusters.
Existing systems achieve this through either intra-operator fusion (IntraFusion), which packs operators into a single large kernel, or inter-operator decomposition (InterDecom), which splits a tensor into multiple parts for pipelined execution. 
However, current IntraFusion methods underutilize network topology, causing suboptimal bandwidth usage on multi-GPU systems, while InterDecom struggles to determine the optimal number of decomposed parts for peak performance. 
To address these issues, we introduce Compass, which employs systematic optimization and comprehensive modeling. First, we design a novel IntraFusion algorithm leveraging double-ring communications to maximize bandwidth utilization in hybrid NVLink-PCIe systems, achieving 1.5x-2.5x speedups. Second, we develop a decomposition model that mathematically derives the optimal tensor decomposition degree for InterDecom, improving performance by up to 1.3x. Finally, we develop a unified performance framework that accurately determines the best strategy for different scenarios.
We validate Compass through extensive evaluation across 288 configurations and end-to-end experiments on real-world applications. The results demonstrate that Compass consistently selects the optimal strategy, achieving up to a 1.42x end-to-end speedup compared to the Megatron-LM baseline. 
\end{abstract}


\section{Introduction}
\label{sec:introduction}

The rapid growth of large language models (LLMs)~\cite{brown2020language, chowdhery2022palm, touvron2023llama} has created severe communication bottlenecks in distributed training, where data exchange latency is now a primary obstacle to computational efficiency~\cite{narayanan2021efficient, canziani2016analysis}. To manage the immense size of these models, various parallelization strategies are employed, including data, pipeline, and tensor parallelism~\cite{li2020pytorch, huang2019gpipe, shoeybi2019megatron}. Among these, tensor parallelism (TP) is particularly critical, as it partitions individual layers across devices to reduce the memory of both model parameters and input activations. However, this strategy introduces intense communication demands. During both forward and backward passes, TP requires frequent collective operations (e.g., \texttt{AllGather} and \texttt{ReduceScatter}) to synchronize the results for each layer. This creates a tight, fine-grained dependency between computation and communication, making latency hiding exceptionally difficult. The frequent synchronization points mean that, in a naive sequential execution, high-performance GPUs are often left stalled, waiting for data from other devices. This communication overhead is a dominant performance bottleneck in large-scale training, often accounting for 40-75\% of the total execution time~\cite{pati2023computation}. Therefore, developing effective techniques to hide this communication latency specifically within TP is challenging.

Two main strategies have been developed to overlap communication with computation. The core principle for both is to break down large communication operators into smaller partitions and interleave their execution with computation tasks. The first approach, \textbf{1) Intra-Operator Fusion} (IntraFusion), packs communication and computation into a single operator. By reordering computation tiles within the General Matrix Multiplication (\texttt{GEMM}) operator, it creates a tight pipeline that hides communication latency without decomposing the operator, thereby maintaining high computational efficiency~\cite{flux2024Chang,comet2025Zhang,coconet2022Jangda}. In contrast, the second approach, \textbf{2) Inter-Operator Decomposition} (InterDecom), pipelines the entire operator by decomposing it into $d$ independent sub-operators. Its theoretical potential for hiding latency is determined by the decomposition granularity, $d$. This approach is more flexible, allowing each sub-task to use highly optimized libraries~\cite{megascale2024Jiang,narayanan2021efficient,wang2022overlap,li2020pytorch}.

However, these methods fail to achieve optimal performance due to three key limitations. First, for IntraFusion, state-of-the-art systems like Flux~\cite{flux2024Chang} are often designed for specific hardware topologies (e.g., fully-connected NVLink topologies), making their communication algorithms inefficient on common hybrid interconnects and sometimes leading to performance even worse than a simple sequential baseline. Second, for InterDecom, performance is highly sensitive to the decomposition degree, $d$, yet current methods lack analytical models to determine the optimal value, forcing reliance on suboptimal heuristics or empirical tuning. Finally, because the two strategies are orthogonal and have different strengths, the best choice depends on both the workload and the underlying hardware, but no existing framework can dynamically select the superior strategy to achieve maximum efficiency.

In this paper, we address these limitations with three key contributions: 1) We design Topology-Aware IntraFusion (\textbf{TA-IntraFusion}) operators for hybrid NVLink-PCIe topologies, using single and double ring algorithms that achieve higher bandwidth utilization than existing methods like Flux~\cite{flux2024Chang}. By reordering \texttt{GEMM} tiling computation to align with ring-based communication patterns, we fully hide communication latency and significantly improve performance over standard sequential implementations. 2) We introduce the Optimal Decomposition Predictor for InterDecom (\textbf{ODP-InterDecom}) that employs an overhead-aware performance model to analytically determine the optimal decomposition degree. Based on this model, we reduce the prediction error by more than 4$\times$ compared to overhead-agnostic models, which eliminates heuristic tuning and improves performance over default configurations. 3) Finally, we build \textbf{Compass}, an intelligent system that automatically selects the optimal communication-hiding strategy. To achieve this, we first construct a unified performance framework by developing a novel performance model for the IntraFusion strategy and integrating it with our InterDecom model. This framework allows Compass to use a lightweight selection algorithm to dynamically determine the superior approach for any given workload and topology.

We validated our approach through an extensive evaluation on a server with 8 NVIDIA A6000 GPUs, spanning 288 distinct test cases. The results validate the effectiveness of each component: TA-IntraFusion achieves operator-level speedups of 1.5$\times$-2.5$\times$ over sequential baselines, while ODP-InterDecom improves performance by up to 1.3$\times$ by replacing suboptimal heuristics. When integrated into Compass, these optimizations yield significant real-world gains. Our Compass consistently selects the optimal strategy, culminating in up to a 1.42$\times$ end-to-end speedup on large-scale models over the highly optimized Megatron-LM baseline. 

\section{Background and Motivations}
\label{sec:background}

\subsection{Communication and Computation in Tensor Parallelism}

\begin{figure}[!t]
    \centering
    \includegraphics[width=\columnwidth]{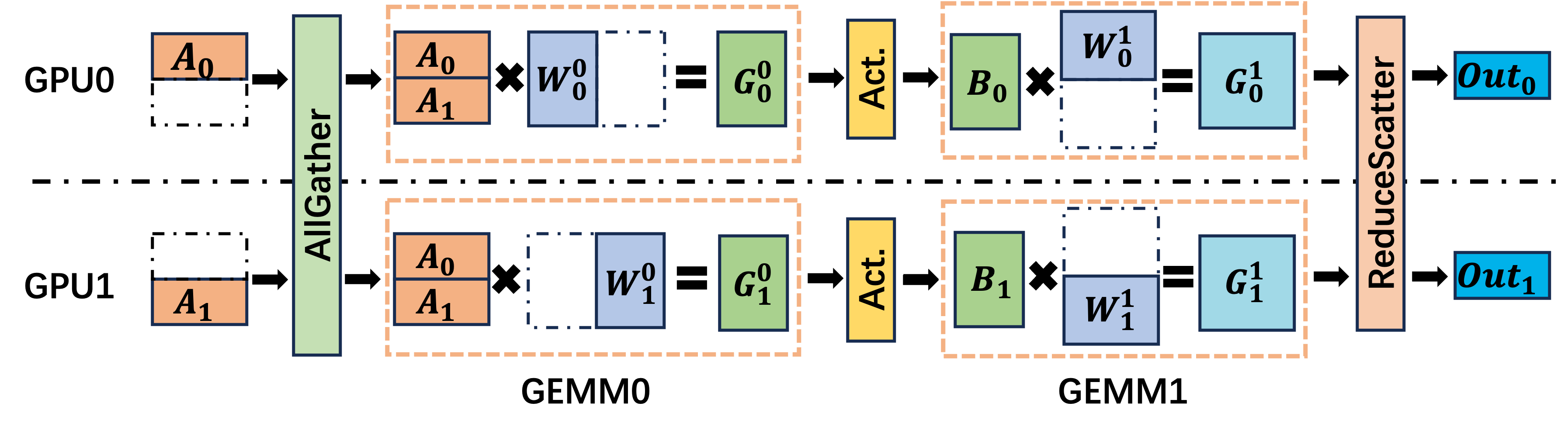}
    \caption{An illustration of a two-layer MLP with tensor parallelism across two GPUs.}
    \label{fig:mlp_fusion}
\end{figure}

Tensor parallelism (TP) is widely used for training large language models~\cite{brown2020language, touvron2023llama}. By partitioning parameters and activations across devices, it reduces memory overhead but introduces expensive collective communications~\cite{korthikanti2023reducing}. As illustrated in Figure~\ref{fig:mlp_fusion}, a standard TP implementation in a Multi-Layer Perceptron (MLP) block consists of a sequence of two GEMM computation operators and two collective communication operators. Specifically, the first weight matrix ($W^0$) of the \texttt{GEMM0} operator is column-parallel partitioned, and the second weight matrix ($W^1$) of the \texttt{GEMM1} operator is row-parallel partitioned. To maintain equivalent output computation results, a preceding \texttt{AllGather} collective is required to obtain the full input before the \texttt{GEMM0} operator (known as \texttt{AllGather+GEMM}), and a subsequent \texttt{ReduceScatter} collective is needed to accumulate and scatter the output of the \texttt{GEMM1} operator (known as \texttt{GEMM+ReduceScatter}).

However, the communication introduced in TP can become a system bottleneck, especially on bandwidth-constrained hardware. In a naive implementation, computation and communication are executed serially: a computational operator must wait for all input from a preceding collective (in \texttt{AllGather+GEMM}), and a collective must wait for all output from a preceding operator before it can begin (in \texttt{GEMM+ReduceScatter}). The sequential execution results in GPU under-utilization and significantly increases end-to-end latency. The frequent intra-layer communication inherent in TP further exacerbates this issue, making communication the primary bottleneck~\cite{pati2023computation}.

\subsection{Computation-Communication Overlapping Strategies}

\begin{figure}[!t]
\centering
\includegraphics[width=\columnwidth]{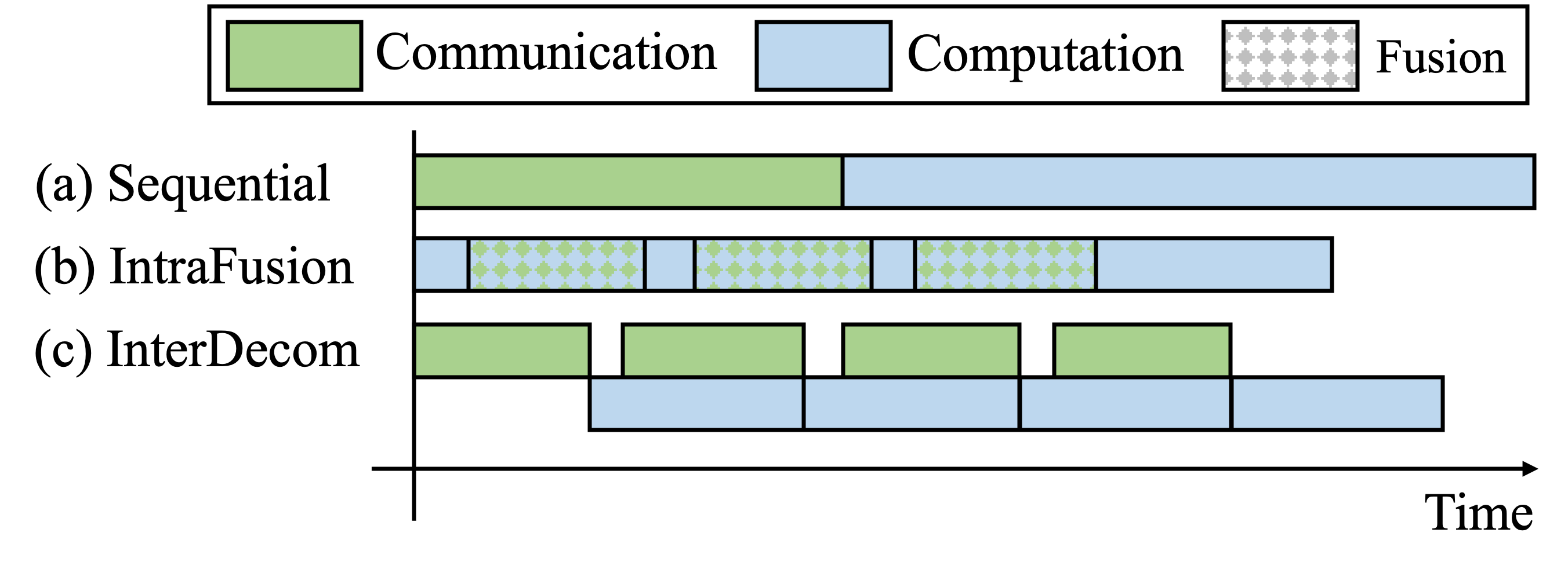}
\caption{Execution timeline of the \texttt{AllGather+GEMM} pattern for three strategies: (a) Sequential execution with no overlap, (b) IntraFusion, and (c) InterDecom.}
\label{fig:fusion_strategies}
\end{figure}

To mitigate the communication bottleneck in TP, modern distributed frameworks employ computation-communication overlapping techniques. The core principle is to break down the atomic communication operator into smaller partitions and interleave their execution with that of adjacent computation tasks, effectively hiding communication latency~\cite{shi2020quantitative,shi2021towards,narayanan2021efficient,wang2022overlap,lamypoirier2023breadthfirst,shah2023taccl}. 
Based on this principle, two fundamental optimization strategies have emerged to implement overlapping in practice, as illustrated in Figure~\ref{fig:fusion_strategies}.

\textbf{Intra-Operator Fusion (IntraFusion)}, illustrated in Figure~\ref{fig:fusion_strategies}(b), overlaps communication with computation by strategically reordering the processing of internal computation tiles within GEMM~\cite{flux2024Chang,comet2025Zhang,coconet2022Jangda}. This reordering ensures that communication latency is concealed. For example, in an \texttt{AllGather+GEMM} pattern, the operator begins processing immediately with locally available data to hide initial transfers, while in a \texttt{GEMM+ReduceScatter} pattern, it schedules the final local computation to overlap with ongoing transmissions. IntraFusion integrates both communication and computation into a single fused operator, eliminating explicit kernel decomposition and minimizing kernel launch overhead.

\textbf{Inter-Operator Decomposition (InterDecom)} achieves overlap by partitioning both computation and communication operators into $d$ smaller, independent sub-operators that can be pipelined~\cite{megascale2024Jiang,narayanan2021efficient,wang2022overlap,harlap2018pipedream}. As shown in Figure~\ref{fig:fusion_strategies}(c), the decomposition degree, $d$, directly controls the pipeline depth and, consequently, its theoretical potential for concealing communication latency. However, since each sub-operator relies on standard collective communication primitives, some communication remains exposed, particularly during the pipeline’s startup and drain phases. This approach offers a flexible method for hiding latency by interleaving at the sub-operator level and supports the use of best-in-class libraries for computation and communication. Sequential execution is a special case where $d=1$.

\subsection{Limitations and Motivations}
\label{subsec:motivation}

Although both IntraFusion and InterDecom aim to hide communication latency, they represent orthogonal approaches, each with significant practical challenges. Crucially, there is no principled guidance to determine which strategy will perform better for a given workload, creating a difficult optimization problem. This section details the inherent limitations of each approach, highlighting the need for a more systematic solution.

\subsubsection{IntraFusion Limitation}

To evaluate IntraFusion, we tested Flux~\cite{flux2024Chang} on a server with 8x Nvidia RTX A6000 GPUs, featuring a hybrid interconnect where all GPUs are connected via PCIe, with some pairs connected via NVLink. Using Flux's benchmark suite\footnote{\url{https://github.com/bytedance/flux/tree/v1.0.4}}, we found that its fusion strategy was often outperformed by a sequential baseline (PyTorch+NCCL (NVIDIA Collective Communications Library)~\cite{li2020pytorch,nvidia2024nccl}). This contrasts sharply with the significant speedups reported in the Flux repository on A800 and H800 systems, which feature fully-connected NVLink topologies. A performance breakdown, detailed in Table~\ref{tab:flux_vs_baseline_detail}, revealed that on our hardware, while computation times were comparable, Flux's communication overhead was substantially higher, transforming the intended overlap into a new bottleneck.

\begin{table}[htbp]
\centering
\caption{Performance breakdown for Sequential vs. Flux \texttt{GEMM+ReduceScatter} (\texttt{GEMM+RS}) and \texttt{AllGather+GEMM} (\texttt{AG+GEMM}) latency on A6000 system with 4 and 8 GPUs. All time are averaged across GPUs and reported in milliseconds (ms).}
\label{tab:flux_vs_baseline_detail}
\begingroup
\renewcommand{\arraystretch}{0.9} 
\begin{tabular}{@{}cllccc@{}}
\toprule
\textbf{TP} & \textbf{Pattern} & \textbf{Strategy} & \textbf{Comp.} & \textbf{Comm.} & \textbf{Total} \\
\midrule
\multirow{4}{*}{4} & \texttt{AG+GEMM} & Sequential & 10.80 & 2.09 & 12.89 \\
& & Flux & 11.16 & 5.80 & 16.96 \\
\cmidrule(l){2-6}
& \texttt{GEMM+RS} & Sequential & 10.92 & 2.49 & 13.40 \\
& & Flux & 11.22 & 6.58 & 17.80 \\
\midrule
\multirow{4}{*}{8} & \texttt{AG+GEMM} & Sequential & 4.88 & 5.50 & 10.38 \\
& & Flux & 5.43 & 12.01 & 17.45 \\
\cmidrule(l){2-6}
& \texttt{GEMM+RS} & Sequential & 4.96 & 6.91 & 11.87 \\
& & Flux & 5.62 & 12.22 & 17.84 \\
\bottomrule
\end{tabular}%
\endgroup
\end{table}

\begin{figure}[!t]
    \centering
    \includegraphics[width=\columnwidth]{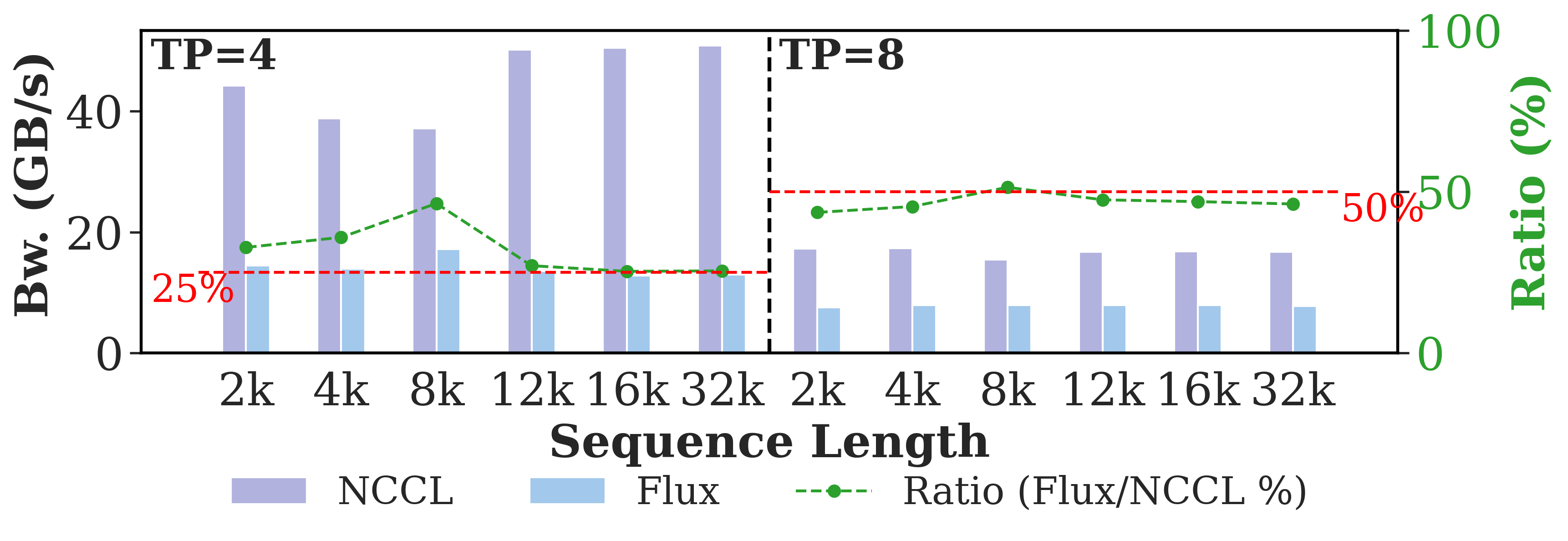}
    \caption{Effective bandwidth (Bw.) comparison of \texttt{AllGather} between NCCL and Flux across different TP sizes. The ratio indicates Flux's performance over NCCL.}
    \label{fig:flux_comm_breakdown}
\end{figure}

To investigate this issue further, we measured the communication performance of Flux’s custom \texttt{AllGather} and compared it with the standard NCCL implementation. The results, shown in Figure~\ref{fig:flux_comm_breakdown}, reveal a significant performance gap: on a 4-GPU setup, Flux achieves only 25\% of NCCL’s effective bandwidth, and on an 8-GPU setup, only 50\%. The root cause is Flux’s reliance on an \texttt{AlltoAll} communication algorithm. While effective on fully connected NVLink topologies, this approach is ill-suited for hybrid interconnects~\cite{shah2023taccl}. On such systems, the \texttt{AlltoAll} pattern, which requires every GPU to communicate simultaneously with all others, inevitably congests lower-bandwidth cross-NUMA links (e.g., QPI). This confirms that current IntraFusion methods, such as Flux, overlook topology complexity and fail to efficiently utilize hybrid interconnect bandwidth. This limitation drives the need to design a topology-aware IntraFusion approach that adapts its communication algorithm to diverse hardware. 

\subsubsection{InterDecom Limitation}

The performance of InterDecom is critically determined by its decomposition degree \(d\). While the strategy aims to hide latency by pipelining \(d\) sub-operators, their concurrent execution leads to resource contention for shared GPU resources like memory bandwidth and multiprocessors~\cite{lee2025characterizing}. This contention introduces a significant overhead. As a result, a larger decomposition degree does not necessarily yield better performance, creating a difficult tuning dilemma. 

\begin{figure}[!t]
\centering
\includegraphics[width=\columnwidth]{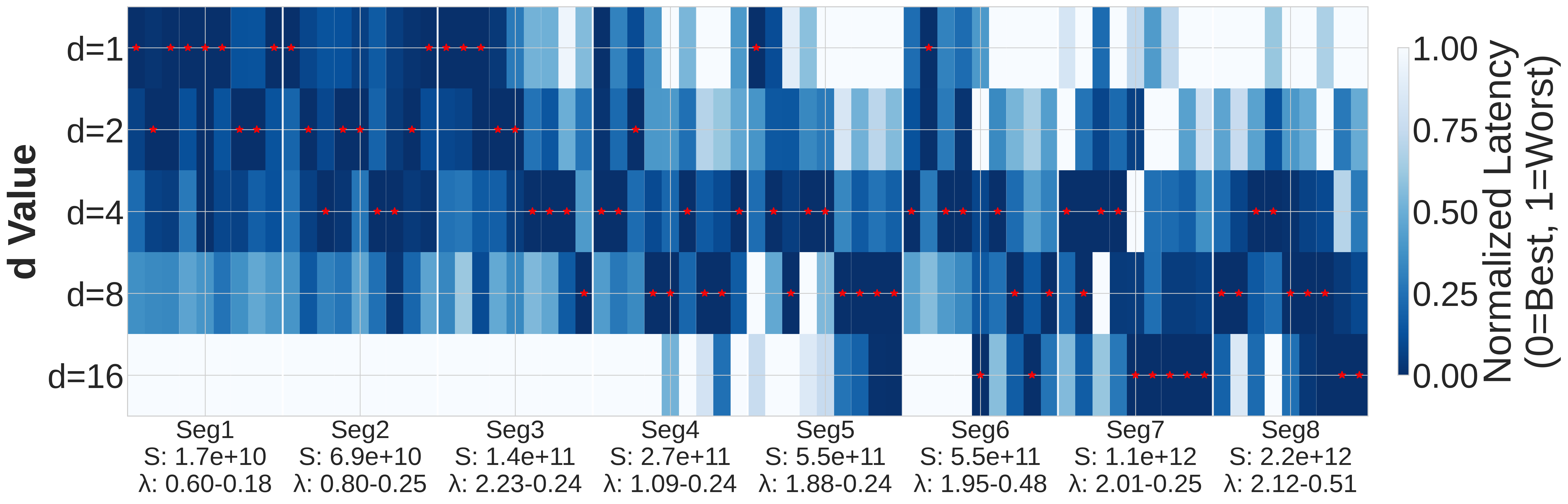}
\caption{The performance landscape of InterDecom across 72 workloads, segmented by problem size ($S$) and computation-to-communication ratio ($\lambda$). Each row corresponds to a different decomposition degree ($d$). The red dot in each column marks the measured optimal $d$ for that specific workload.}
\label{fig:optimal_k_distribution}
\end{figure}

Figure~\ref{fig:optimal_k_distribution} vividly illustrates this tuning dilemma by visualizing performance across 72 workloads. To isolate the impact of $d$ while controlling for factors such as operator type and parallelism degree, it displays performance for the \texttt{AllGather+GEMM} pattern at TP=8 across all workloads listed from Table~\ref{tab:experimental_config}. The color of each cell indicates performance relative to the best choice for that workload (dark blue is best, white is worst). The landscape is treacherous: a suboptimal choice of $d$ incurs a severe penalty, averaging 57.44\% and reaching up to 281.90\%. The lack of principled guidance for selecting $d$ exacerbates this issue, forcing practitioners to rely on empirical tuning or simple heuristics. Such methods are unreliable; for instance, the common default of $d=4$~\cite{megascale2024Jiang} is optimal in only 30\% of cases tested. Consequently, achieving optimal performance with the InterDecom strategy is challenging, often leading to significantly suboptimal outcomes. This underscores the critical need for a predictor to accurately determine the optimal decomposition degree. 

\subsubsection{The Strategy Selection Limitation}

One fundamental limitation of current overlapping methods is that the optimal choice between the two orthogonal strategies (IntraFusion and InterDecom) depends heavily on the specific workload and hardware, yet no existing framework can make this selection automatically. To quantify this challenge, we conducted an extensive benchmark of the \texttt{AllGather+GEMM} pattern across 144 diverse workloads (Table~\ref{tab:experimental_config}). The results revealed no definitive winner: IntraFusion was superior in 78 cases (54\%), while InterDecom excelled in the remaining 66 (46\%). This finding empirically confirms that the best-performing strategy is highly context-dependent, underscoring the critical need for a principled method to guide this choice.

\section{TA-IntraFusion: Topology-Aware IntraFusion}
\label{sec:tile_scheduling_optimization}

\subsection{Topology-Aware Communication Algorithm}
\label{subsec:topology_aware_comm}

Our topology-aware algorithm resolves the communication bottleneck of existing fusion strategies by matching communication patterns to physical hardware, thereby maximizing the bandwidth utilization of hybrid interconnects.

\begin{figure}[!t]
    \centering
    \includegraphics[width=\columnwidth]{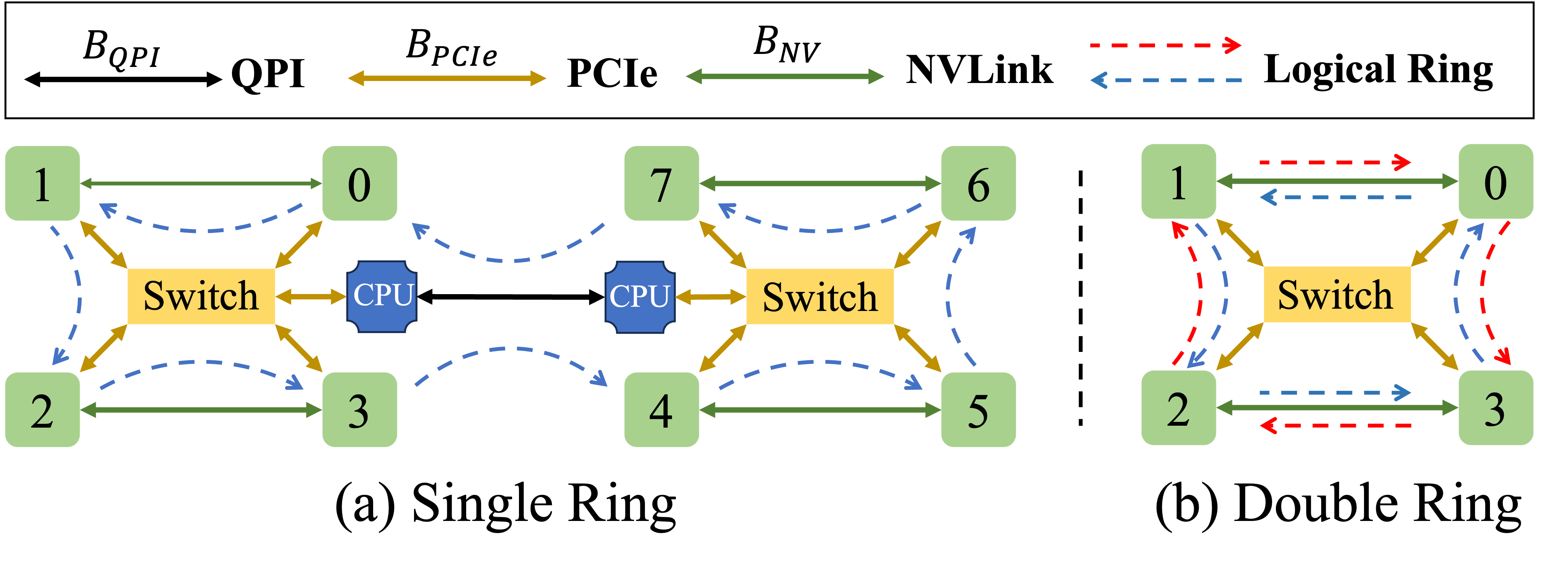}
    \caption{Single Ring (a) and Double Ring (b) communication algorithms on hybrid interconnects. (a) An 8-GPU single ring with cross-NUMA. (b) A 4-GPU double ring with concurrent forward (blue) and backward (red) rings.}
    \label{fig:double_ring_topology}
\end{figure}

In scenarios requiring cross-NUMA (Non-Uniform Memory Access) communication, we utilize a topology-aware single ring algorithm. As depicted in Figure~\ref{fig:double_ring_topology}(a), this strategy connects all GPUs into a logical ring to perform communication in serialized steps. Crucially, only two links (0--7 and 3--4) traverse the NUMA boundary, which significantly alleviates pressure on the higher-level interconnect (e.g., QPI). Assuming each of the 8 GPUs holds data of size \(m\), the communication proceeds in 7 steps, each transferring a chunk of size \(m/8\). The performance is thus bottlenecked by the slowest link in the ring (the QPI, with bandwidth \(B_{\text{QPI}}\)), resulting in a total effective bandwidth of approximately \((8/7)B_{\text{QPI}}\). In contrast, a standard \texttt{AlltoAll} algorithm, used in Flux, requires all four GPUs on one NUMA domain to simultaneously communicate with the four on the other. This heavily congests the QPI link, reducing the effective bandwidth to approximately \(B_{\text{QPI}}/2\), about half that of our single-ring approach. This confirms our experimental findings in Figure~\ref{fig:flux_comm_breakdown}.

Within a single NUMA domain, we employ a double ring on hybrid topologies with both PCIe and NVLink interconnects~\cite{nvidia2024rtxa6000, huawei2024atlasa3}. As illustrated in Figure~\ref{fig:double_ring_topology}(b), this algorithm establishes two concurrent, counter-rotating rings (e.g., a forward and a backward ring). By carefully mapping these logical rings to the physical interconnects, each GPU can simultaneously use both its PCIe and NVLink interfaces. For instance, communication between GPUs 0--1 and 2--3 is routed over NVLink, while connections between GPUs 1--2 and 0--3 use PCIe. In this manner, the duplex channels of each hardware interconnect are fully utilized. Because data is transferred on both rings in parallel, the effective bandwidth becomes \((8/3)B_{\text{PCIe}}\), where \(B_{\text{PCIe}}\) is the PCIe's bandwidth. This delivers a 2x performance improvement compared to a single ring's \((4/3)B_{\text{PCIe}}\) bandwidth and a nearly 4x improvement over a standard \texttt{AlltoAll} algorithm on this topology. This confirms our experimental observations in Figure~\ref{fig:flux_comm_breakdown}.

\subsection{Mechanics of TA-IntraFusion}
\label{subsec:fusion_mechanics}

We implement both single ring and double ring communication algorithms using dedicated Direct Memory Access (DMA) engines and a push-based model to maximize asynchronous overlap~\cite{jena2019gpudirect, sccl2021cai}. This is then tightly integrated with a customized CUTLASS~\cite{cutlass2017} \texttt{GEMM} kernel that prioritizes tile processing based on data readiness and uses atomic semaphores for exquisite synchronization. This combination of enhancements produces fusion operators where the computation schedule adapts to data readiness, allowing the operator to fully hide communication latency while exploiting the system's hybrid interconnect.

\begin{figure}[!t]
    \centering
    \includegraphics[width=\columnwidth]{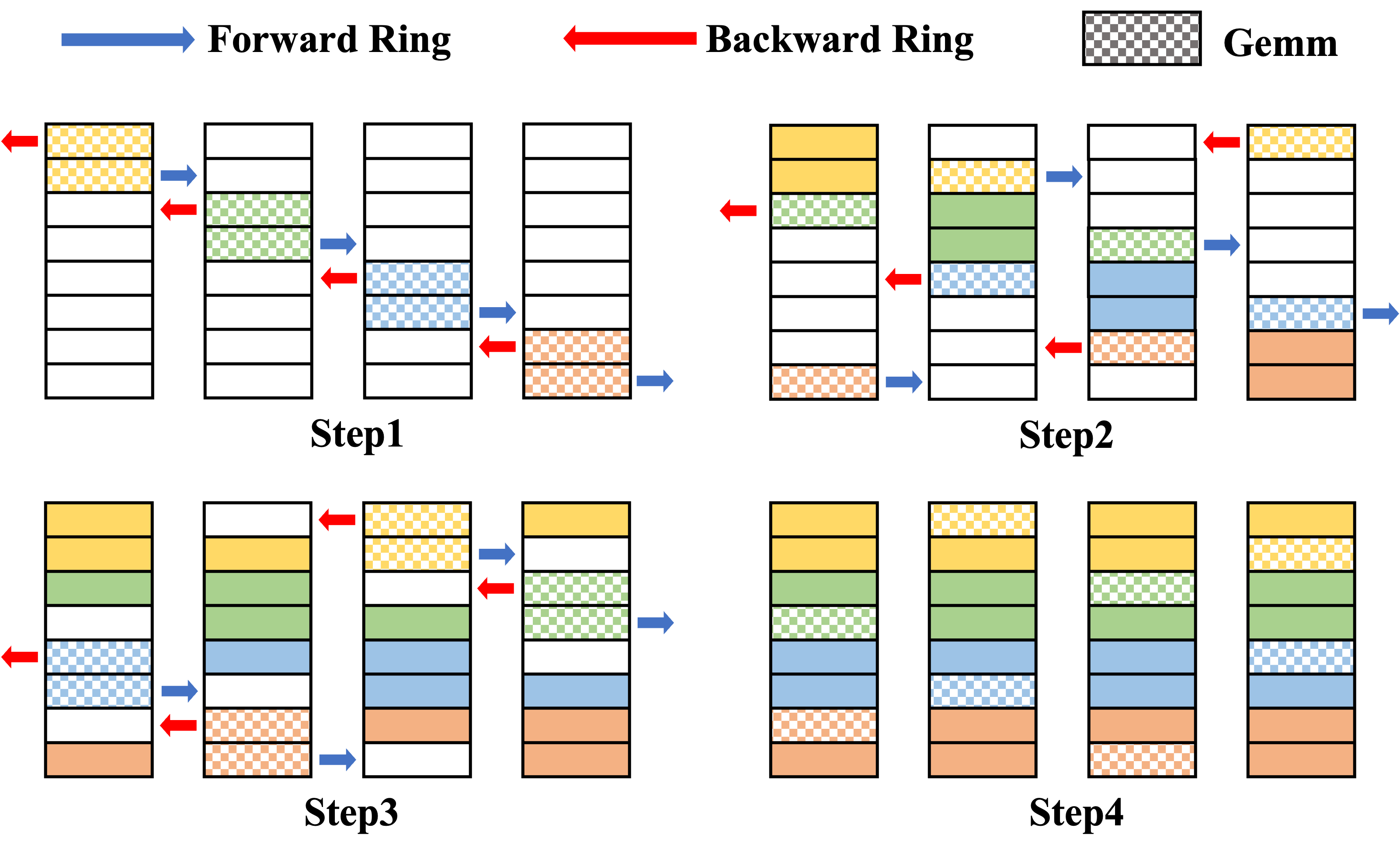}
    \caption{Example of \texttt{AllGather+GEMM} fusion on a 4-GPU double ring.}
    \label{fig:double_ring_fusion}
    \vspace{-1em}
\end{figure}

Figure~\ref{fig:double_ring_fusion} illustrates the mechanics of our double ring fusion with a 4-GPU \texttt{AllGather+GEMM} example. The mechanics of a single ring are analogous but simpler, involving data transfer in only one direction. Initially, each GPU holds its local data. To align with the communication algorithm for four GPUs, the final output tensor of the \texttt{AllGather} is first conceptually partitioned into four chunks. Each of these chunks is then further subdivided into a forward ring and a backward ring component, resulting in eight logical partitions.

The execution flow is designed to maximize computation-communication overlap. As shown in Step 1 in Figure~\ref{fig:double_ring_fusion}, the process commences with each GPU performing \texttt{GEMM} computation (indicated by the checkerboard pattern) on its local data, which is immediately available. This initial computation hides the latency of the first data transfers. Concurrently, the double ring communication is initiated: the backward ring (red arrows) circulates even-indexed partitions, while the forward ring (blue arrows) circulates odd-indexed partitions in the opposite direction. This orchestrated, bidirectional data flow creates a ``rolling wave'' of data availability. In subsequent steps, the arrival of a remote data partition at a GPU immediately triggers the corresponding \texttt{GEMM} computation for that partition. This pipelined process continues until all partitions are gathered and computed, ensuring that communication and computation are performed in parallel throughout the operator, thus minimizing overall execution time.

For brevity, we omit the \texttt{GEMM+ReduceScatter} case. The principle is analogous, but the dependency is reversed: communication relies on computation. To maximize overlap, each GPU strategically schedules the computation for the data partition that will remain on the local device to be last. This allows the GPU to overlap its local computation with the ongoing transmission of data from other GPUs, effectively hiding communication latency.

\section{ODP-InterDecom: Optimal Degree Predictor}
\label{sec:pipeline_scheduling_optimization}

\subsection{Problem Formulation}

The primary challenge in InterDecom is selecting the optimal decomposition degree, $d$. This choice involves a critical trade-off: a smaller $d$ restricts overlap potential, whereas a larger $d$ increases overlap but incurs higher overhead~\cite{lee2025characterizing}, resulting in diminishing returns. 

The goal is to find the optimal degree $d^*$ that minimizes the total execution time. This can be formally expressed as:
\begin{equation}
d^* = \arg\min_{d} T_{decomp}(\mathcal{W}, \mathcal{T}, d),
\end{equation}
\noindent where $\mathcal{W}$ and $\mathcal{T}$ represent the workload characteristics and hardware topology, respectively. To solve this, we must first establish an accurate, overhead-aware model for $T_{decomp}$.

\subsection{Decomposition Execution Time Model}
\label{subsec:pipeline_scheduling_optimization_execution_time_model}

To predict the performance of InterDecom, we model the total execution time, $T_{decomp}(\mathcal{W}, \mathcal{T}, d)$, with the following formula:
\begin{align}
T_{decomp}(\mathcal{W}, \mathcal{T}, d) &= \underbrace{(d-1) \cdot T_{stage} \cdot \eta_{overhead}^{decomp}(d)}_{\text{Overlapped stages}} \label{eq:pipeline_model} \\
             &\quad + \underbrace{T_{comp}^{(1)} + T_{comm}^{(1)}.}_{\text{Non-overlapped portion}} \nonumber
\end{align}

This model consists of two primary components: the overlapped stages subject to overhead, and a final non-overlapped portion.

Each of the $(d-1)$ overlapped stages takes $T_{stage} = \max(T_{comp}/d, T_{comm}/d)$. Following prior work~\cite{shi2023pipemoe}, we model the base computation ($T_{comp}$) and communication ($T_{comm}$) times as linear functions of workload ($\mathcal{W}$) and hardware ($\mathcal{T}$) parameters.

Prior overhead-agnostic models, such as that in FasterMoE~\cite{he2022fastermoe}, fail to account for the significant performance degradation caused by real-world resource contention. As shown in Table~\ref{tab:pipe_model_accuracy_comparison}, this omission leads to substantial prediction errors that grow with the decomposition degree, $d$. Upon further investigation of the per-stage overhead factor, we found that it exhibits a clear linear relationship with $d$. Therefore, we introduce an overhead factor, $\eta_{overhead}^{decomp}(d)$, and model it as a linear function of $d$ to accurately capture this effect:

\begin{equation}
\eta_{overhead}^{decomp}(d) = \alpha_{decomp} + \beta_{decomp} \cdot d,
\label{eq:pipeline_overhead_model}
\end{equation}
\noindent where $\alpha_{decomp}$ represents the baseline coordination overhead and $\beta_{decomp}$ captures the incremental overhead per stage. When this overhead is set to 1, our model degenerates to the simpler overhead-agnostic model. 

Finally, the non-overlapped portion represents the pipeline startup and drain phases that cannot be overlapped. This consists of the time to process the first and last sub-operator, which is the sum of their computation and communication times, $T_{comp}^{(1)} + T_{comm}^{(1)} = T_{comp}/d + T_{comm}/d$.

This overhead-aware model provides significantly more accurate prediction of actual execution times compared to prior approaches that ignore overhead. As validated by the results in Table~\ref{tab:pipe_model_accuracy_comparison}, incorporating the overhead factor dramatically enhances the model's predictive power, reducing the overall Mean Absolute Error (MAE)~\cite{willmott2005advantages} from 5.43 ms to just 0.96 ms. This substantial improvement confirms that our overhead-aware model successfully captures the critical performance dynamics of decomposition execution, providing a reliable foundation for the analytical optimization that follows.

\begin{table}[htbp]
    \centering
    \caption{Comparison of Prediction Accuracy Between Overhead-agnostic and Overhead-aware InterDecom Models.}
    \label{tab:pipe_model_accuracy_comparison}
    \begingroup
    \setlength{\tabcolsep}{4pt} 
    \begin{tabular}{@{}lccccc@{}}
    \toprule
    & \multicolumn{5}{c}{\textbf{Decomposition Degree ($d$)}} \\
    \cmidrule(lr){2-6}
    \textbf{Metric} & \textbf{2} & \textbf{4} & \textbf{8} & \textbf{16} & \textbf{Overall} \\
    \midrule
    \multicolumn{6}{l}{\textit{Model Prediction MAE (ms)}} \\
    \addlinespace
    \hspace{1em}Overhead-agnostic & 3.51 & 4.69 & 5.59 & 7.94 & \textbf{5.43} \\
    \hspace{1em}Overhead-aware  & 0.91 & 1.39 & 1.01 & 0.55 & \textbf{0.96} \\
    \midrule
    \textbf{Avg. Execution Time (ms)} & 25.44 & 24.45 & 24.40 & 26.27 & \textbf{25.14} \\
    \bottomrule
    \end{tabular}
    \endgroup
\end{table}

\subsection{Optimal Decomposition Degree Selection}
\label{subsec:pipeline_scheduling_optimization_optimal_decomposition}

Having established an overhead-aware execution model, we now determine the optimal decomposition degree $d^*$ that minimizes total execution time. The execution time model, $T_{decomp}(d)$, is a strictly convex function with respect to the decomposition degree $d$. Within a specific regime (computation- or communication-bound), its mathematical form simplifies to $f(d) = Ad + B + C/d$, where the coefficients $A$ and $C$ are positive constants. The second derivative, $f''(d) = 2C/d^3$, is always positive for $d>0$, confirming the function's convexity. This property guarantees that a unique minimum exists, which can be found by setting the first derivative to zero. This analytical optimization is applied for the domain $d>1$. For the sequential case where $d=1$, the overlapped stage term in Equation~\ref{eq:pipeline_model} vanishes, and the model correctly yields the execution time $T_{comp} + T_{comm}$. The candidate degree, $d_{cand}^*$, found through this derivation must then be explicitly compared against this baseline to ensure a net performance gain. The final optimal degree, $d_{final}^*$, is thus determined by the following selection logic:
\begin{equation}
\label{eq:optimal_d_final}
d_{final}^* = 
\begin{cases} 
    d_{cand}^* & \text{if } T_{decomp}(d_{cand}^*) < T_{decomp}(1) \\ 
    1 & \text{otherwise},
\end{cases}
\end{equation}
\noindent where the candidate degree $d_{cand}^*$ is given by:
\begin{equation}
\label{eq:d_cand_formula}
d_{cand}^* = \mathcal{P}_2\left(\sqrt{\frac{1 - \alpha_{decomp} + 1/\lambda}{\beta_{decomp}}}\right),
\end{equation}
where, the function $\mathcal{P}_2(\cdot)$ rounds its argument to the nearest power of two. The coefficients $\alpha_{decomp}$ and $\beta_{decomp}$ are determined via a one-time, offline linear regression, while the computation-to-communication ratio, $\lambda = T_{comp} / T_{comm}$, is calculated online from the workload characteristics. This ensures we only apply decomposition when it yields a clear performance advantage.

The model also reveals several intuitive trade-offs. First, $d^*$ is inversely proportional to $\lambda$; for compute-bound workloads (large $\lambda$), a smaller $d$ is preferred to maintain high GEMM kernel efficiency by minimizing fragmentation. Conversely, as the overhead coefficients $\alpha_{decomp}$ and $\beta_{decomp}$ increase, indicating a higher cost for overlapping, the optimal $d$ naturally decreases to favor a less aggressive decomposition strategy.

\section{Compass: Optimal Strategy Selector}
\label{sec:modeling_and_selector}

\subsection{Problem Formulation}

As demonstrated in Section~\ref{sec:background}, neither IntraFusion nor InterDecom is universally superior. This creates a critical challenge: how to automatically select the strategy that minimizes latency for a given workload $\mathcal{W}$ and hardware topology $\mathcal{T}$.

The complete optimization problem can be formally expressed as selecting the strategy with the minimum predicted execution time:
\begin{equation}
\text{Strategy}^* = \arg\min \{T_{\text{fusion}}(\mathcal{W}, \mathcal{T}), T_{\text{decomp}}(\mathcal{W}, \mathcal{T}, d^*)\}.
\end{equation}
\noindent To solve this, we must compare the predicted time of the fusion strategy against that of the decomposition strategy operating at its optimal degree, $d^*$. This optimal degree is itself determined using the analytical solution derived in Section~\ref{subsec:pipeline_scheduling_optimization_optimal_decomposition}. Since the model for $T_{decomp}$ has been detailed in Section~\ref{sec:pipeline_scheduling_optimization}, we next focus on developing an equally robust, overhead-aware model for $T_{fusion}$.

\subsection{IntraFusion Model}
\label{subsec:fusion_overhead_model}

IntraFusion represents a fundamentally different approach from InterDecom. Decomposition achieves overlap by partitioning computation and communication operators into $d$ separate sub-operators and pipelining their execution to achieve a parallelism degree of $d$. In contrast, fusion maintains the operator's integrity by reordering its internal tile-based computation schedule and invoking carefully orchestrated asynchronous communication APIs within the single operator. This distinction leads to entirely different sources of overhead: for decomposition, the primary cost arises from resource contention between its computation and communication sub-operators, whereas for fusion, the overhead stems from the synchronization required to coordinate its internal logical chunks. Consequently, a model tailored for decomposition's resource contention is not suited for fusion, necessitating a distinct analytical model to capture its unique performance characteristics.

The core of a fusion operator, as illustrated in Figure~\ref{fig:fusion_strategies}(b), involves dividing the workload into $n_{chunks}$ logical steps. To match the communication pattern of ring-based algorithms, $n_{chunks}$ is typically set to the number of GPUs. In our double ring strategy, each of these logical chunks is further partitioned into forward-ring and backward-ring components. This structural efficiency allows us to formulate the total execution time as follows:
\begin{align}
\label{eq:fusion_latency_model}
T_{fusion}(\mathcal{W}, \mathcal{T}) &= \sum_{i=2}^{n_{chunks}} \max(T_{comp}^{chunk(i)}, T_{comm}^{chunk(i)}) \nonumber \\
&\quad + T_{comp}^{chunk_{1}} + T_{overhead}^{fusion}.
\end{align}

The model is composed of three distinct components: 1) The summation term represents the main overlapped execution phase across $n_{chunks}-1$ logical pipeline stages. 2) Because communication is effectively hidden by computation during these $n_{chunks}-1$ steps, only the computation for a single logical chunk remains exposed, denoted by the $T_{comp}^{chunk_{1}}$ term. 3) The $T_{overhead}^{fusion}$ term captures the additional overhead from internal synchronization, which is quantified below.

The overhead term $T_{overhead}^{fusion}$ quantifies the coordination cost at the $n_{chunks}-1$ synchronization points during overlapped execution. This overhead is pattern-dependent and scales with the message size due to the need to synchronize metadata across all participating nodes. We model this overhead as:
\begin{equation}
T_{overhead}^{fusion} = (n_{chunks}-1) \cdot \left(\alpha_{fusion} + \beta_{fusion} \cdot m\right),
\end{equation}
\noindent where $\alpha_{fusion}$ captures the base synchronization latency, $\beta_{fusion}$ represents the scaling factor for message-size-dependent coordination costs, and $m$ denotes the communication volume. For \texttt{AllGather} operator, $m = M \cdot K$, while for \texttt{ReduceScatter} operator, $m = M \cdot N$.

The accuracy of our overhead-aware model is validated against an overhead-agnostic baseline~\cite{he2022fastermoe}. Our model demonstrates a dramatic improvement in prediction accuracy, reducing the Mean Absolute Error (MAE) by over 4x, from 10.87 ms to just 2.66 ms. This substantial gain confirms that ignoring coordination overhead leads to poor predictions and validates that our additive overhead model successfully captures the essential performance dynamics of fusion-based strategies, providing a reliable foundation for the strategy selector.



\subsection{Adaptive Strategy Selection Algorithm}

Having established validated performance models for each execution strategy, we now present an adaptive algorithm that automatically selects the optimal strategy for a given workload-topology pair. This model-driven approach eliminates the need for heuristic thresholds by leveraging analytical models to make performance-optimal decisions. As detailed in Algorithm~\ref{alg:compass_selector}, our approach operates in three distinct phases: 1) it first computes the base performance metrics for the given workload and topology; 2) it then loads pre-calibrated overhead parameters and analytically determines the optimal decomposition degree $d^*$; and finally, 3) it uses the validated models and derived parameters to predict the execution time for both strategies and selects the superior one. The algorithm's time complexity is $O(1)$ since all performance predictions involve closed-form calculations, making it suitable for runtime decision-making in production environments.

\begin{algorithm}[htbp]
    \caption{Compass Strategy Selection Algorithm}
    \label{alg:compass_selector}
    \begin{algorithmic}[1]
    \Require Workload parameters $\mathcal{W}$, Topology $\mathcal{T}$
    \Ensure Optimal strategy $S^*$ and predicted execution time $T^*$
    
    \Function{SelectOptimalStrategy}{$\mathcal{W}, \mathcal{T}$}
        \State \textbf{// Phase 1: Compute Base Performance Metrics}
        \State $T_{\text{comp}} \gets f(\mathcal{W}, \mathcal{T})$
        \State $T_{\text{comm}} \gets g(\mathcal{W}, \mathcal{T})$
        \State $\lambda \gets T_{\text{comp}} / T_{\text{comm}}$ \Comment{Comp-to-comm ratio}
        \State \textbf{// Phase 2: Parameter and Degree Calculation}
        \State $(\alpha_{\text{decomp}}, \beta_{\text{decomp}}) \gets$ GetCalibratedParameters$()$
        \State $(\alpha_{\text{fusion}}, \beta_{\text{fusion}}) \gets$ GetCalibratedParameters$()$
        \State $d^* \gets$ GetOptimalDecompDegree$()$
        \State \textbf{// Phase 3: Prediction and Strategy Selection}
        \State $T_{\text{decomp}} \gets$ GetDecompTime$(\mathcal{W}, \mathcal{T}, d^*)$ \Comment{Eq.~\ref{eq:pipeline_model}}
        \State $T_{\text{fusion}} \gets$ GetFusionTime$(\mathcal{W}, \mathcal{T})$ \Comment{Eq.~\ref{eq:fusion_latency_model}}
        \State $T^* \gets \min(T_{\text{fusion}}, T_{\text{decomp}})$
        \State $S^* \gets \begin{cases}
            \text{TA-IntraFusion}, & \text{if } T_{\text{fusion}} < T_{\text{decomp}} \\
            \text{ODP-InterDecom}, & \text{otherwise}
        \end{cases}$
        \State \textbf{return} $(S^*, T^*)$
    \EndFunction
\end{algorithmic}
\end{algorithm}

\section{Evaluation}
\label{sec:evaluation}

This section presents an empirical evaluation of our work. We first validate the performance of our individual optimizations, namely the TA-IntraFusion and the ODP-InterDecom. We then demonstrate the accuracy and end-to-end effectiveness of the Compass selection system.

\subsection{Experimental Setup}
All experiments were conducted on a server with the hardware configuration detailed in Table~\ref{tab:server-config}. The software stack includes Ubuntu 22.04 LTS, CUDA Toolkit 12.6.2, PyTorch 2.7.0, NCCL 2.27.3, CUTLASS 3.9.0, and is compiled with GCC 12.5. Our experimental test cases span a comprehensive parameter space that covers representative distributed deep learning scenarios~\cite{narayanan2021efficient}. Table~\ref{tab:experimental_config} details the configuration parameters used in our evaluation, which collectively generate 288 distinct test cases. This extensive parameter sweep enables systematic analysis of strategy selection patterns across diverse workload characteristics and system scales.

\begin{table}[!t]
    \centering
   
      \caption{Testbed Hardware Specifications.}
     \label{tab:server-config}
   
   \begin{tabular}{@{}ll@{}}
   \toprule
   \textbf{Component}    & \textbf{Specification} \\ 
   \midrule
   CPU     & Intel® Xeon® Platinum 8358, @2.60 GHz \\
   GPU     & 8x Nvidia RTX A6000, 48GB GDDR6 \\
   NVlink  & 112.5GB/s (4x) \\ 
   PCIe    & 4.0 (x16) \\ 
   \bottomrule
\end{tabular}
\end{table}

\begin{table}[!t]
\centering
\caption{Experimental Configuration Parameters.}
\label{tab:experimental_config}
\begin{tabular}{@{}ll@{}}
\toprule
\textbf{Parameter} & \textbf{Candidate Values} \\
\midrule
Hidden Dim. (K) & $\{4096, 8192, 16384\}$ \\
Intermediate Dim. (N) & $\{2048, 4096, 8192, 16384\}$ \\
World Size (GPUs) & $\{4, 8\}$ \\
Decomposition Degree & $\{2, 4, 8, 16\}$ \\
Sequence Length (M) & $\{2048, 4096, 8192, 12288, 16384, 32768\}$ \\
Fusion Pattern & $\{$\texttt{AG+GEMM}, \texttt{GEMM+RS}$\}$ \\
\bottomrule
\end{tabular}
\end{table}

\subsection{Effectiveness of Optimizations}
\label{sec:effectiveness_of_optimizations}

\subsubsection{Operator-Level Performance Comparison of TA-IntraFusion}

\begin{figure}[!t]
    \centering
    \begin{subfigure}[b]{\columnwidth}
        \centering
        \includegraphics[width=\textwidth]{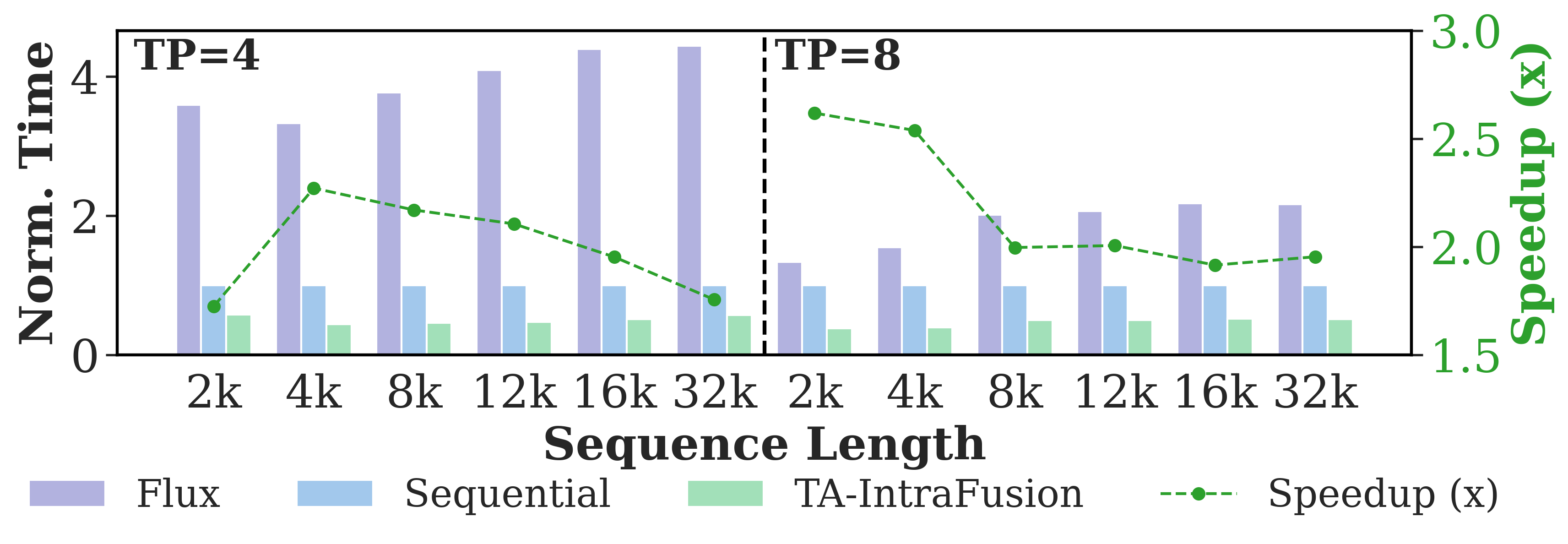}
        \caption{\texttt{AllGather+GEMM} Performance Comparison.}
        \label{fig:ag_perf_comp}
    \end{subfigure}
    
    \vspace{0.3cm}
    
    \begin{subfigure}[b]{\columnwidth}
        \centering
        \includegraphics[width=\textwidth]{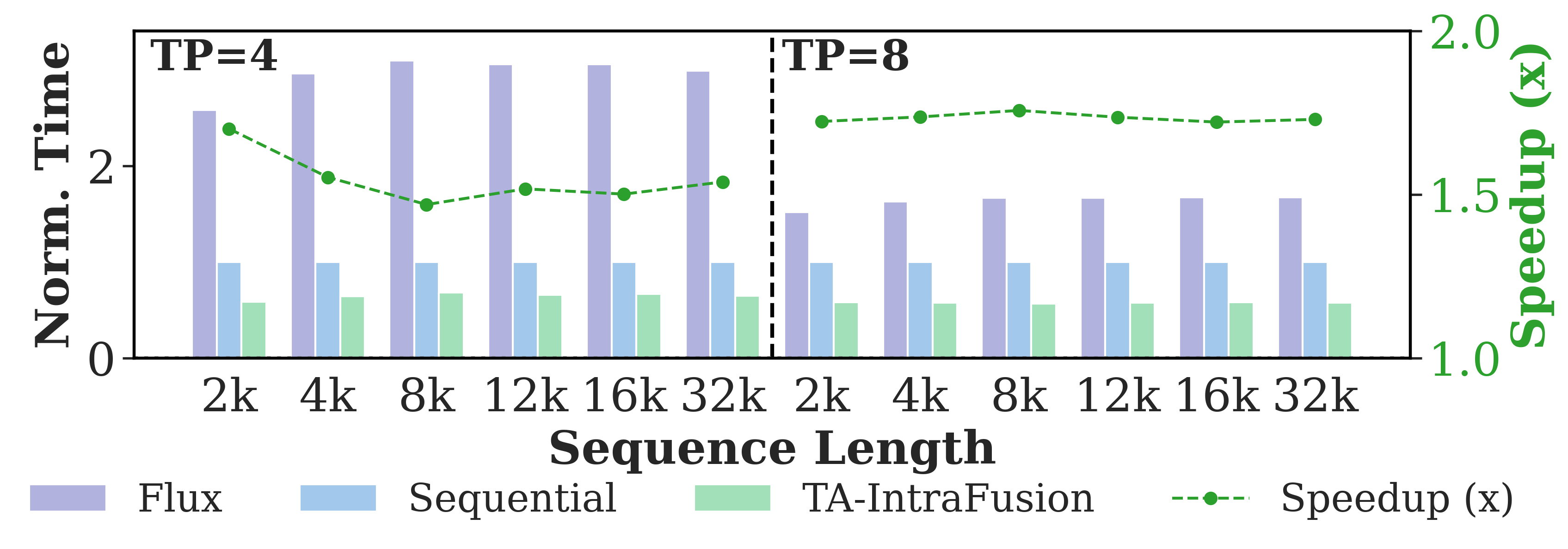}
        \caption{\texttt{GEMM+ReduceScatter} Performance Comparison.}
        \label{fig:rs_perf_comp}
    \end{subfigure}
    \caption{Operator-Level performance comparison of our fusion operator against Flux and a Sequential (PyTorch+NCCL) baseline for TP=4 and TP=8. The bar charts show execution time normalized to the sequential baseline. The green line indicates the speedup (x) of our method over the sequential baseline.}
    \label{fig:e2e_fusion_perf}
\end{figure}

To evaluate the operator-level performance of TA-IntraFusion, we selected a representative set of results from the benchmark suite defined in Table~\ref{tab:experimental_config} for visualization. Figure~\ref{fig:e2e_fusion_perf} displays these cases, comparing our method against two critical baselines: Flux~\cite{flux2024Chang} and a standard sequential implementation (PyTorch+NCCL)~\cite{li2020pytorch,nvidia2024nccl}. The charts show performance for both \texttt{AllGather+GEMM} and \texttt{GEMM+ReduceScatter} fusion operators across tensor parallelism sizes of 4 and 8.

The results, presented in Figure~\ref{fig:e2e_fusion_perf}, consistently demonstrate the superiority of our TA-IntraFusion operator. Our method significantly outperforms both the topology-agnostic Flux baseline and the standard sequential implementation. For the \texttt{AllGather+GEMM} pattern (Figure~\ref{fig:e2e_fusion_perf}(a)), we achieve speedups ranging from 1.5x to 2.5x. This indicates that communication is not only effectively hidden but, in many cases, the fused operation is faster than the baseline communication itself. This is because our operator leverages a high-bandwidth DMA copy engine, which surpasses the performance of the Streaming Multiprocessor (SM) copy mechanism used by the default NCCL baseline~\cite{shah2023taccl}. For the \texttt{GEMM+ReduceScatter} pattern (Figure~\ref{fig:e2e_fusion_perf}(b)), speedups are typically between 1.5x and 2.0x. The slightly lower, yet still substantial, speedup is due to minor resource competition between the \texttt{GEMM} and the \texttt{reduce} computations that must run concurrently within the fusion operator, which slightly impacts the overall efficiency of the overlap. These results validate that TA-IntraFusion design is highly effective at hiding communication latency.

\subsubsection{Optimal Decomposition Degree Selection}
\label{sec:eval_pipeline_tuning}

\begin{figure}[!t]
    \centering
    \includegraphics[width=\columnwidth]{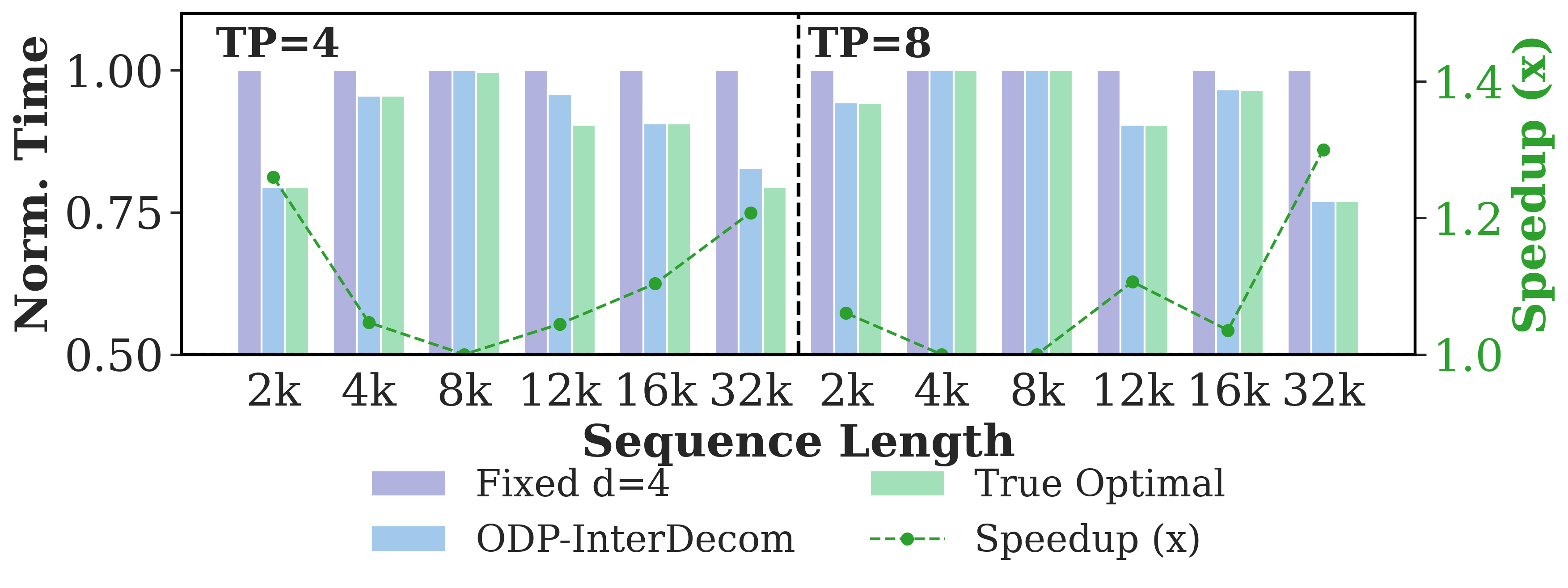}
    \caption{Performance comparison between our predicted optimal decomposition degree, a fixed degree ($d=4$), and the true optimal. The bar charts show execution time normalized to the fixed-degree ($d=4$) baseline, while the green line indicates the speedup (x) of our method over the same baseline.}
    \label{fig:pipeline_degree_comparison}
\end{figure}

This section evaluates the practical benefits of our analytical approach for selecting the optimal pipeline decomposition degree, $d^*$. We compare the performance of our method, which dynamically predicts the optimal $d$ for each workload, against two key baselines: a common heuristic that uses a fixed decomposition degree ($d=4$) and the empirically measured true optimal performance. The results, presented in Figure~\ref{fig:pipeline_degree_comparison}, show the normalized execution time for all three methods and the corresponding speedup achieved by our approach over the fixed-degree baseline across various sequence lengths for TP sizes of 4 and 8.

The results, which hold across all 288 test cases defined in Table~\ref{tab:experimental_config}, underscore a critical insight: the optimal decomposition degree, $d^*$, is not static but varies with workload parameters such as sequence length and tensor parallelism size. Consequently, a common heuristic using a fixed degree (e.g., $d=4$) fails to adapt, leading to suboptimal performance. In contrast, our analytical model dynamically identifies the optimal $d^*$ for each workload. As shown in Figure~\ref{fig:pipeline_degree_comparison}, this adaptive approach not only consistently outperforms the fixed-degree baseline but also achieves performance nearly identical to the true optimal, validating the high accuracy of our model. The performance gap widens as sequence length increases, with our method delivering speedups of over 1.2x for TP=4 and nearly 1.3x for TP=8 at the largest sequence lengths. When the model correctly predicts $d^*=4$, its performance matches the baseline, further confirming the model's accuracy.

\subsubsection{End-to-End Performance Evaluation}
\label{subsec:e2e_performance_evaluation}

\begin{figure}[!t]
    \centering
    \begin{subfigure}[b]{\columnwidth}
        \centering
        \includegraphics[width=\textwidth]{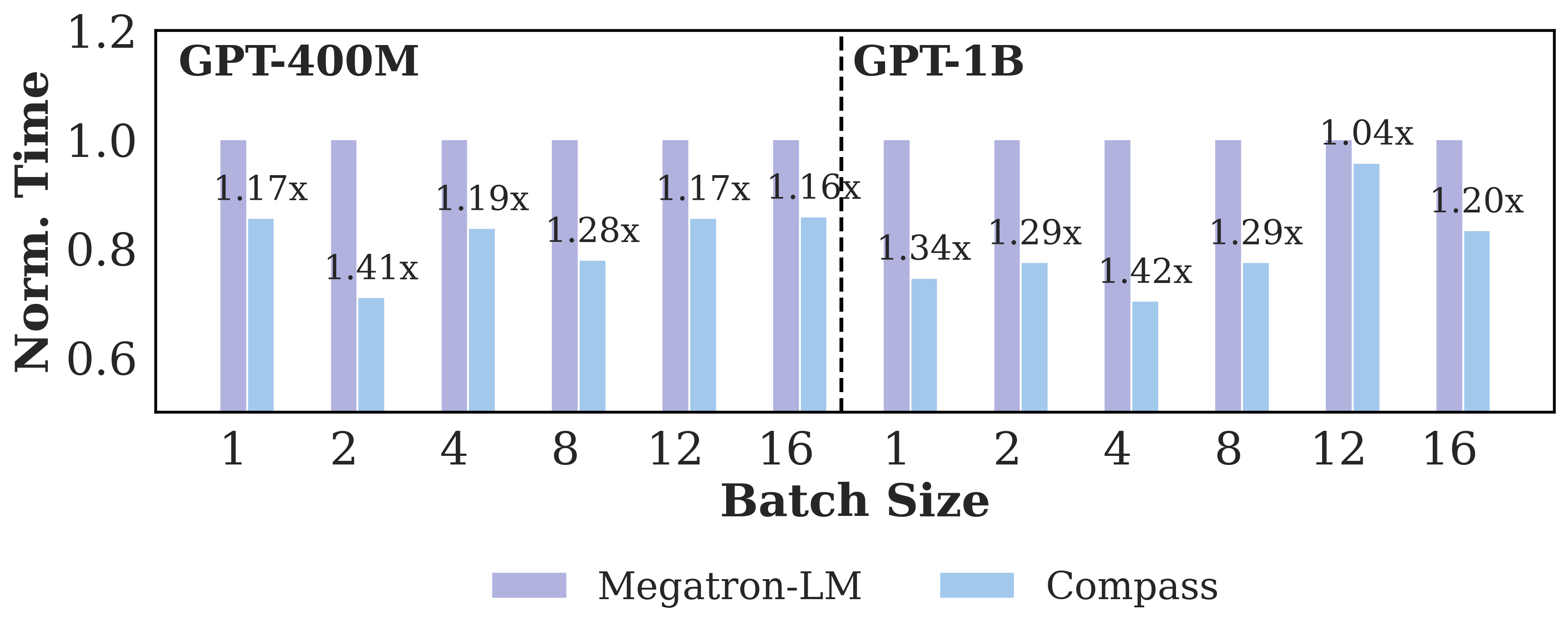}
        \caption{End-to-end performance comparison on GPT models.}
        \label{fig:e2e_gpt}
    \end{subfigure}
    
    \vspace{0.3cm}
    
    \begin{subfigure}[b]{\columnwidth}
        \centering
        \includegraphics[width=\textwidth]{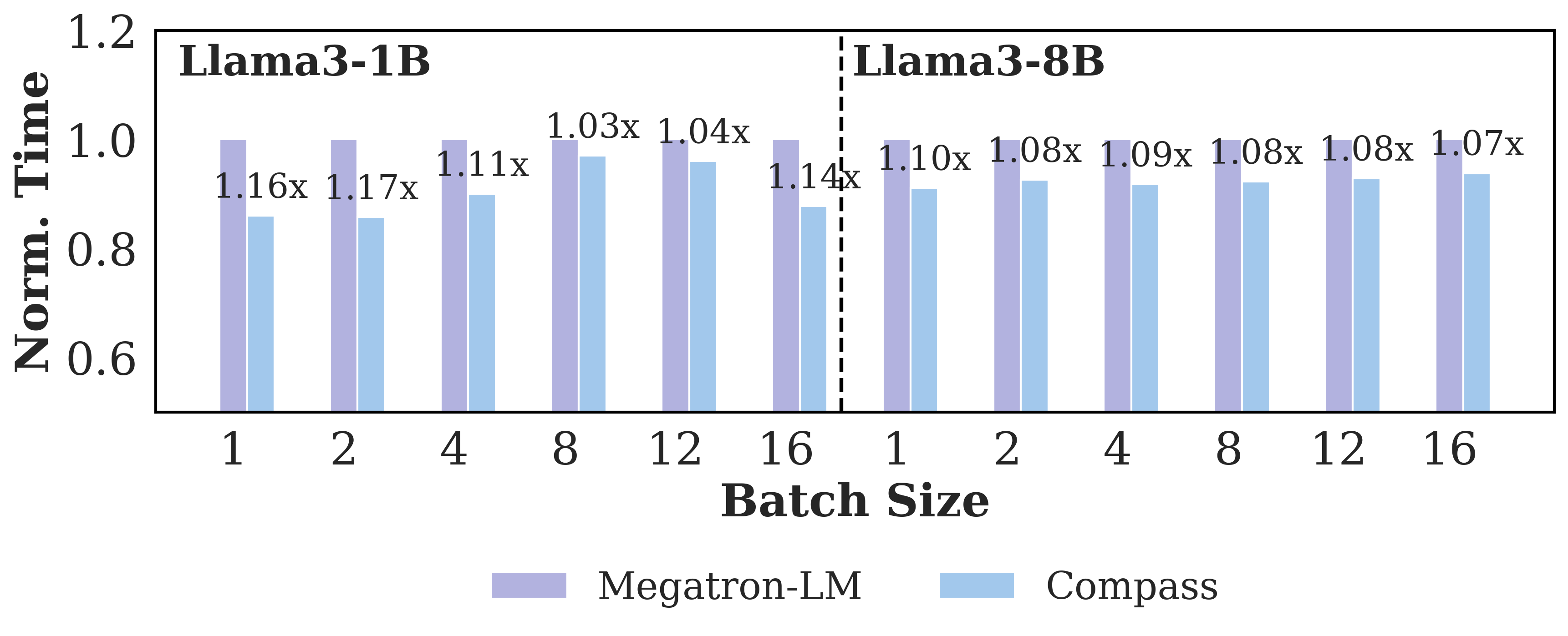}
        \caption{End-to-end performance comparison on Llama3 models.}
        \label{fig:e2e_llama}
    \end{subfigure}
    \caption{End-to-end performance comparison between Compass and Megatron-LM on GPT and Llama3 models. }
    \label{fig:e2e_performance_comparison}
\end{figure}

To evaluate the real-world efficacy of our framework, we compared Compass against the widely-used Megatron-LM~\cite{megatron-lm} in end-to-end forward pass experiments on both GPT and Llama3 models. The evaluation was conducted on 8x Nvidia RTX A6000. For GPT models, we used a 400M-parameter version with a TP degree of 4 and a 1B-parameter version with TP=8, following established guidelines~\cite{gptmodelguide}. For Llama3 models, we used a 1B-parameter version of Llama3 with a TP degree of 4 and an 8B-parameter version with TP=8. As shown in Figure~\ref{fig:e2e_performance_comparison}, Compass delivers substantial real-world performance gains, achieving speedups of up to 1.42x on GPT models and 1.17x on Llama3 models over the Megatron-LM baseline.


\subsection{Discussions}
\label{sec:discussions}

\subsubsection{Strategy Selection Performance}
\label{sec:strategy_selection_evaluation}

\begin{figure}[!t]
    \centering
    \includegraphics[width=\columnwidth]{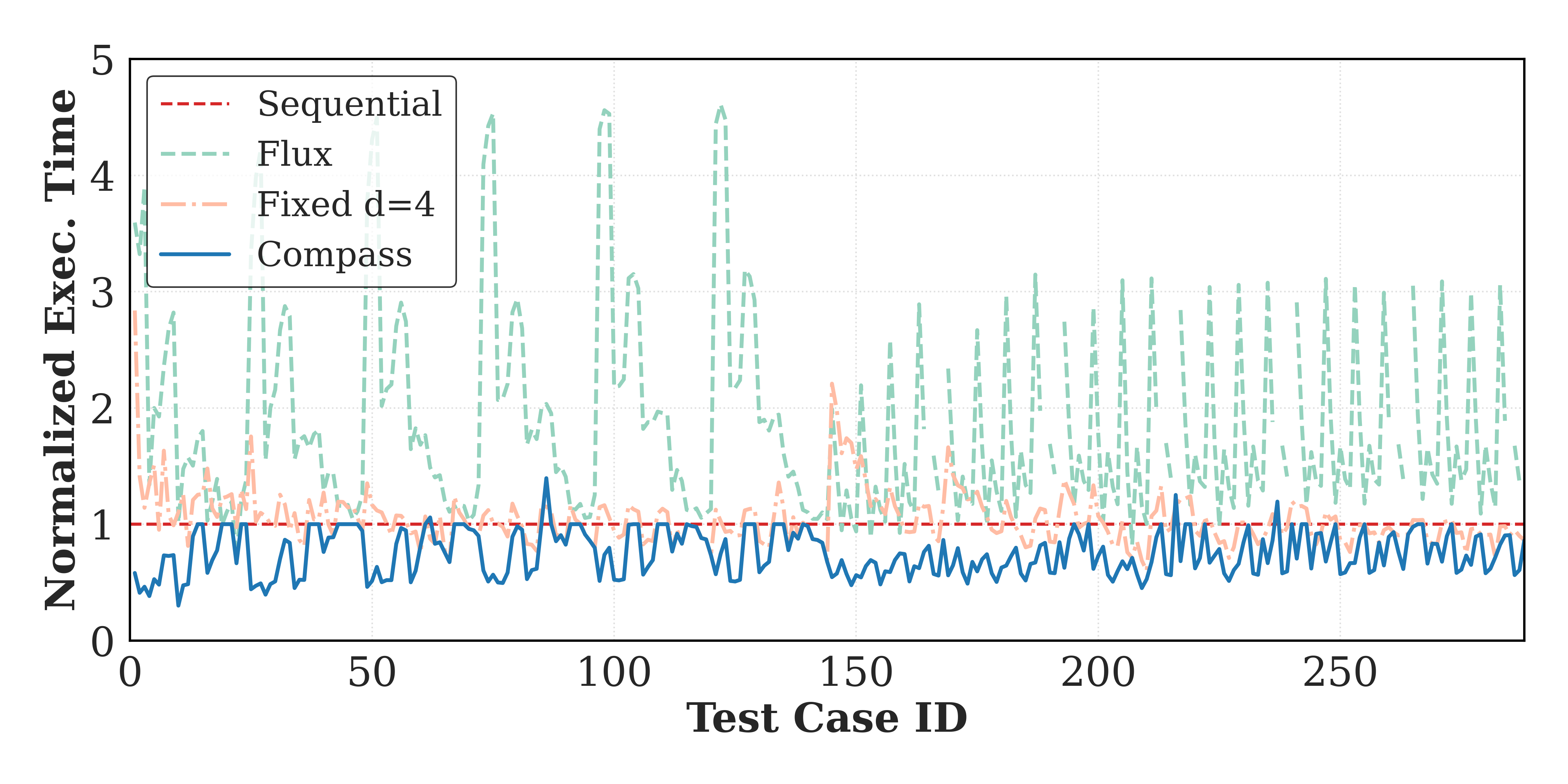}
    \caption{Normalized exec. time comparison.}
    \label{fig:line_compare}
\end{figure}

The tangible performance benefits of our complete strategy selector confirm the central thesis of this paper: no single strategy is universally optimal. As shown in Figure~\ref{fig:line_compare}, Compass consistently delivers the best performance, successfully navigating the treacherous performance landscape of fixed strategies. For example, in certain cases, choosing an optimal strategy can yield a 2x speedup over the sequential baseline, while the worst-case one can make the execution 5 times slower. Our selector achieves an overall accuracy of 92.67\% in choosing the optimal strategy across 288 test cases, and this high accuracy is the foundation of the end-to-end speedups observed in Figure~\ref{fig:e2e_performance_comparison}. Notably, the results also reveal that the optimal strategy exhibits different patterns for \texttt{AllGather} and \texttt{ReduceScatter} workloads, a phenomenon we plan to investigate in future work.

\subsubsection{TA-IntraFusion Communication Micro-benchmark}
\label{sec:comm_microbenchmark}

\begin{figure}[!t]
    \centering
    \includegraphics[width=\columnwidth]{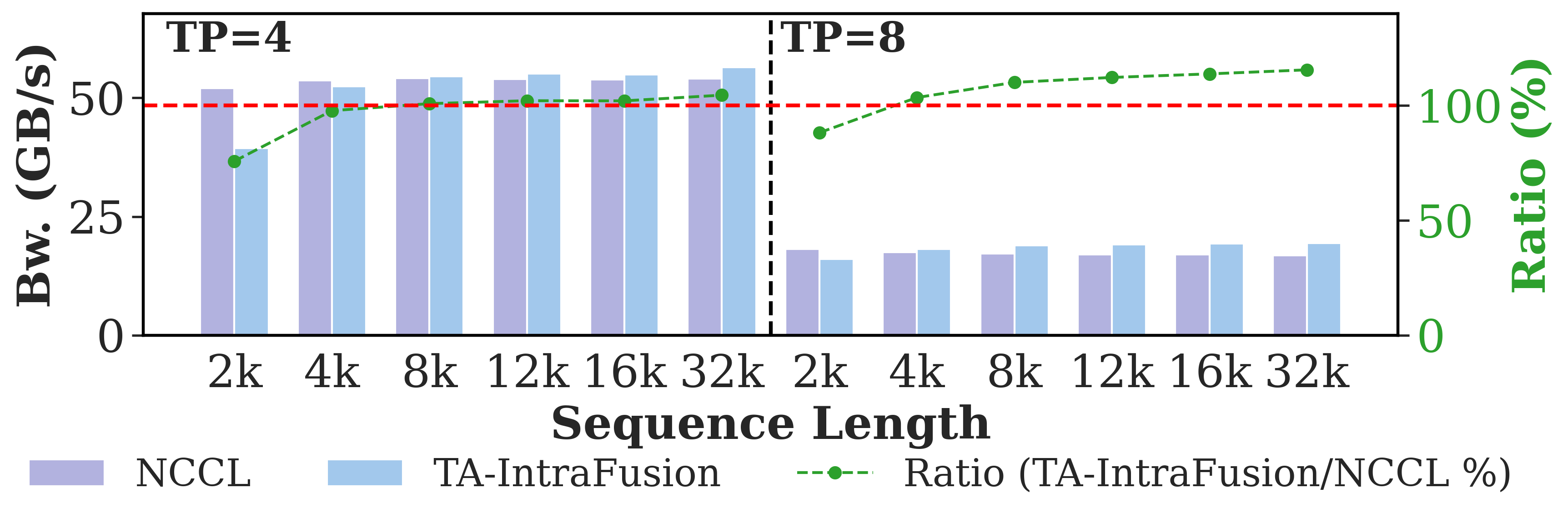}
    \caption{Communication performance benchmark of our TA-IntraFusion operator against NCCL.}
    \label{fig:comm_breakdown}
\end{figure}

To validate the communication performance of our TA-IntraFusion operator, we conducted a micro-benchmark that isolates its data transfer component and compares its effective bandwidth against the highly-optimized NCCL library. The results, presented in Figure~\ref{fig:comm_breakdown}, show that our fusion operators achieve bandwidth comparable to, and often surpassing, the NCCL baseline as message size grows. Notably, for the smallest message sizes, NCCL shows a slight advantage. This is attributable to the lower startup overhead of its default SM Copy Mechanism compared to our DMA Copy Engine approach. However, as the data volume increases, the higher bandwidth utilization of the dedicated DMA Copy Engine becomes dominant, allowing our operator to consistently outperform the baseline~\cite{sccl2021cai}. The ratio line rising above 100\% confirms we have eliminated the communication bottleneck that plagued prior fusion works like Flux, validating that our communication design is robust and highly optimized.

\section{Related Work}
\label{sec:related}

This section presents related work on communication optimizations in distributed deep learning.

Efficient communication primitives are vital for accelerating large language model training. While early ring-based algorithms were common, they struggle at scale. Consequently, tree-based approaches~\cite{nvidia2024nccl} and hybrid strategies~\cite{patarasuk2009bandwidth, bienz2019nodeaware} have emerged. These hybrid methods optimize for heterogeneous interconnects by decomposing communication into topology-aware stages.  

For tensor-parallel communication overlapping, works typically use two strategies. IntraFusion (e.g., Flux~\cite{flux2024Chang}, Comet~\cite{comet2025Zhang}, CoCoNet~\cite{coconet2022Jangda}) reorders computations within a single operator but often suffers from topology-agnostic designs. InterDecom pipelines sub-operators, as seen in Megascale~\cite{megascale2024Jiang} and others~\cite{wang2022overlap, huang2019gpipe, narayanan2019pipedream}, but relies on heuristically tuned decomposition degrees. While both strategies have been explored independently, to the best of our knowledge, no existing work provides a systematic method to model both and automatically select the optimal strategy. 

Beyond single-layer communication scheduling, system-level approaches optimize training efficiency through task scheduling. ACP-WFBP~\cite{shi2021exploiting}  schedules tasks for tensor fusion~\cite{sergeev2018horovod,shi2019mg,shi2021mg} with simultaneous communications which are better to utilize network bandwidth. DualPipe~\cite{deepseekv3_2024} pipelines model components to minimize communication overhead and boost hardware utilization. Similarly, Centauri~\cite{chen2024centauri} uses global scheduling to dynamically manage tasks and dataflows, enhancing system throughput and efficiency. These optimizations are orthogonal to our work.

\section{Conclusion}
\label{sec:conclusion}


In this paper, we present Compass, a system that employs systematic optimization and comprehensive modeling to achieve optimal computation-communication overlap in large language model training. First, Compass introduces a novel TA-IntraFusion algorithm that enhances bandwidth utilization on hybrid topologies by taking advantage of ring algorithms and aligning computation with communication patterns. Second, it incorporates a highly accurate performance model to determine the optimal decomposition degree, eliminating costly empirical tuning. Third, Compass integrates a unified performance framework to dynamically select the best strategy for any workload and hardware configuration. Our evaluation shows that Compass consistently outperforms existing approaches, achieving up to a 1.42× end-to-end speedup on real-world applications. 

\section*{Acknowledgments}
\label{sec:Acknowledgment}
This work was partially supported by National Natural Science Foundation of China under Grant No. 62272122, the Guangzhou Municipal Joint Funding Project with Universities and Enterprises under Grant No. 2024A03J0616, Guangzhou Municipality Big Data Intelligence Key Lab (2023A03J0012), Hong Kong CRF grants under Grant No. C7004-22G and C6015-23G.

\bibliographystyle{IEEEtran_etal}
\bibliography{references}
\end{document}